\documentclass[aps,preprint,groupedaddress]{revtex4-1}
\usepackage{epsfig,graphicx,amsmath,amssymb}
\usepackage{subfigure}

\begin{document}
\title{Accelerating Solitons in Gas-Filled Hollow-Core Photonic Crystal Fibers}
\author{M.~\surname{Fac\~ao}}
\email{mfacao@ua.pt}
\affiliation{Departamento de F\'{i}sica, Universidade de Aveiro and I3N
Campus Universit\'ario de Santiago, 3810-193 Aveiro, Portugal}
\author{M.~I.~\surname{Carvalho}}
\email{mines@fe.up.pt}
\affiliation{DEEC/FEUP and INESCPorto, Universidade
do Porto, Rua Dr. Roberto Frias, 4200-465 Porto, Portugal}
\author{P.~\surname{Almeida}}
\email{palmeida@ua.pt}
\affiliation{Departamento de Matem\'{a}tica, Universidade de Aveiro,
Campus Universit\'ario de Santiago, 3810-193 Aveiro, Portugal}

\begin{abstract}
We found the self-similar solitary solutions of a recently proposed model for propagation of
pulses in gas filled hollow-core photonic crystal fibers that includes a plasma induced nonlinearity. As anticipated for a simpler model and using a perturbation analysis, there are indeed stationary solitary waves that accelerate and self-shift to higher frequencies. However, if the plasma nonlinearity strength is large or the pulse amplitudes are small, the solutions have distinguished long tails and decay as they propagate.
\end{abstract}
\maketitle

\section{Introduction}

Hollow-core photonic crystal fibers (HC-PCFs) can exhibit very
interesting properties, such as relatively low loss, low group
velocity dispersion and high confinement of light in the core
\cite{russell03,russell06}, while also allowing new nonlinear
phenomena associated with the interaction of light and matter
filling these fibers. Lately, these HC-PCFs have been filled with
gases for purposes of enhancing Raman scattering if a Raman-active
gas is used \cite{benabid02}, or further controlling the total
dispersion of the fiber by varying the gas pressure \cite{nold10}.
Furthermore, it has also been shown that few $\mu$J or even picoJ
\cite{holzer11,husko13} energy optical pulses are sufficient to
ionize the gas and produce a plasma, leading to new nonlinear effects
such as the blueshifting of the central wavelength of the pulses
\cite{fedotov07,holzer11}. Despite the fact that the soliton shift to
higher frequencies has also been reported in other contexts, such
as in a line-defect waveguide \cite{colman12} and in tapered
solid-core photonic crystal fibers \cite{stark11}, the existence of
a blueshift in a Raman-active gas opens new exciting opportunities
of controlling the soliton dynamics by two competing processes, one
leading to a redshift, usually known as soliton self-frequency shift (SSFS) caused
by intrapulse Raman scattering (IRS) \cite{gordon86}, and the other to a blueshift.

Traditionally, the interaction between light and matter has been
studied using computationally demanding methods based on models for the full
electric field of the pulse \cite{geissler99} but, recently, Saleh
\emph{et al.} presented a model that describes pulse propagation in
hollow-core photonic crystal fibers filled with a gas as a pair of
coupled equations for the electric field envelope and ionization
fraction \cite{saleh11}. This model, which neglects losses and
results from a linearization of the tunneling model for pulse
intensities close to the threshold intensity, has proved to be
amenable to the application of both numerical and analytical
techniques. In effect, by using a perturbation approach the
occurrence of the blueshift effect has already been adequately
predicted \cite{saleh11}.

In this work, we present a thorough study of accelerating solitons
in gas-filled HC-PCF, extending the results in \cite{saleh11}. We
start with the model proposed by Saleh \emph{et al.} \cite{saleh11},
use an accelerating self-similarity variable to obtain an ordinary
differential equation (ODE) to which we apply a perturbation
approach and solve using a shooting procedure. In this analysis, we
have considered the exact solution for the ionization term and our
results apply to both zero and nonzero threshold intensities. The
dependence on the model parameters, namely, the plasma and
Raman strengths, the intensity threshold and the pulse peak value
are studied in detail.

\section{Self-similarity variable and perturbation approach}

As mentioned in the introduction, here we will follow Saleh \emph{et al.} \cite{saleh11} and start with the following coupled equations
\begin{equation}
i\frac{\partial\psi}{\partial z}-\frac{\beta_2}{2}\frac{\partial^2
\psi}{\partial t^2}+\gamma|\psi|^2\psi-\gamma t_R(|\psi|^2)_t\psi
-\frac{\omega_p^2}{2k_0c^2}\psi=0
\end{equation}
\begin{equation}
\frac{\partial n_e}{\partial t}=(\tilde{\sigma}/A_\text{eff})(n_T-n_e)\Delta |\psi|^2\Theta(\Delta |\psi|^2)
\label{eq-ne}
\end{equation}
where $\psi(z,t)$ is the optical field envelope in units of square
root of power, $z$ is the distance along the fiber, $t$ is the time
in a reference frame moving with the group velocity at the central
frequency $\omega_0$, $\beta_2$ is the group velocity dispersion
parameter, $\gamma$ is the nonlinear parameter, $t_R$ is the Raman parameter, $c$ is the vacuum speed
of light, $k_0=\omega_0/c$,
$\omega_p=[e^2n_e/(\epsilon_0m_e)]^{1/2}$ is the plasma frequency
associated with an electron density $n_e(t)$, $e$ and $m_e$ are the
electron charge and mass, respectively, $\epsilon_0$ is the
vacuum permitivity and $A_\text{eff}$ is the effective mode area. The
plasma-induced nonlinearity only occurs for intensities above the
threshold intensity $I_\text{th}=|\psi|^2_\text{th}/A_\text{eff}$,
so that $\Delta|\psi|^2=|\psi|^2-|\psi|^2_\text{th}$ and $\Theta$ is
the Heaviside step function. $n_T$ is the total number density of
ionizable atoms, associated with the maximum plasma frequency
$\omega_T=(e^2n_T/(\epsilon_0m_e)]^{1/2}$ and $\tilde{\sigma}$ is
the photoionization cross-section. This model assumes that the
recombination time is longer than the pulse and neglects the
ionization induced loss that is small especially for pulses whose
maximum is barely above the threshold. The equation (\ref{eq-ne})
may be solved exactly and after an adimensionalization we obtain
\begin{widetext}
\begin{equation}
i\frac{\partial q}{\partial\xi}+\frac{1}{2}\frac{\partial^2 q}{\partial \tau^2}
%+q\int_{-\infty}^\tau R(\tau')|q|^2d\tau'
+|q|^2q-\tau_R(|q|^2)_\tau q-\phi_Tq\left(1-\text{e}^{-\sigma\int_{-\infty}^\tau\Delta|q|^2\Theta(\Delta|q|^2)d\tau'}\right)=0
\label{pde}
\end{equation}
\end{widetext}
where $q=(\gamma z_0)^{1/2}\psi$, $\xi=z/z_0$, $\tau=t/t_0$,
$\tau_R=t_R/t_0$,  $\phi_T=\tfrac{1}{2}k_0z_0(\omega_T/\omega_0)^2$,
$\sigma=\tilde{\sigma}t_0/(A_\text{eff}\gamma z_0)$, where
$z_0=t_0^2/|\beta_2|$ is the called the dispersion length and $t_0$
is an arbitrary time chosen similar to the pulse duration. Motivated
by the observation of blueshifting of the pulses and previous
perturbation approaches, we used an accelerating variable
$T=\tau+\frac{a}{4}\xi^2+b\xi$ and solutions of the form
$q(\xi,\tau)=F(T)\exp(i\theta(\xi,T))$, with $F$ and $\theta$ real,
and obtained an ODE for $F$
\begin{widetext}
$$F''+aTF-DF+2F^3-4\tau_R F^2F'-2\phi_TF\left(1-\text{e}^{-\sigma\int_{-\infty}^T\Delta F^2\Theta(\Delta F^2)dT'}\right)=0,$$
\end{widetext}
and the following expression for the phase
\begin{equation}\theta(\xi,T)=-\left(\frac{a}{2}\xi+b\right)T+\frac{1}{2}(D+b^2)\xi+\frac{1}{4}ba\xi^2+\frac{1}{24}a^2\xi^3+E
\label{phase}
\end{equation}
where $D$ and $E$ are arbitrary constants. In order to reduce the
number of parameters, we introduced the following change of variables
$P(\zeta)=\sigma F(T)$ and $T=\sigma \zeta$, with which the ODE for $P(\zeta)$
reads
\begin{widetext}
\begin{equation}
P'' + \alpha \zeta P - CP + 2{P^3} - {\gamma _R}{P^2}P' - {\gamma
_P}P\left( {1 - {{\text{e}}^{ - \int_{ - \infty }^\zeta {\Delta
{P^2}\Theta (\Delta {P^2})d\zeta'} }}} \right) = 0 \label{Peq}
\end{equation}
\end{widetext}
with $\alpha  = a{\sigma ^3}$, ${\gamma _R} = 4\tau _R/\sigma$ and
${\gamma _P} = 2{\phi _T}{\sigma ^2}$. If we further define $${\gamma _P} =
\chi,\quad {\gamma _R} = R\chi$$ where $R =\gamma _R/\gamma _P$, it will permit the application of a
perturbation approach simultaneously to the two terms, namely, Raman and plasma, as long as $\chi$ is small.

Hence, we have used a perturbation approach around the ODE
associated with the nonlinear Schr\"odinger equation (NLSE) whose
results are valuable by themselves if the additional terms are
small, but that also serve as first estimates for our shooting
method. Hence, we consider expansions for $P$ and $\alpha$ in powers
of $\chi$ such that
$$P=G(\zeta-\zeta_0)+\chi P_1(\zeta)+\cdots,$$
where $G(\zeta-\zeta_0)=\sqrt{C}\text{sech}(\sqrt{C}(\zeta-\zeta_0))$ and
$$\alpha=\chi\alpha_1+\cdots$$
and introduce them in (\ref{Peq}). To first order, we obtain
\begin{multline*}
P_1''-CP_1+6G^2P_1=-\alpha_1\zeta G+RG^2G'\\
-G\left(1-\text{e}^{-\int_{-\infty}^\zeta(G^2-P_\text{th}^2)\Theta(G^2-P_\text{th}^2)d\zeta'}\right)
\end{multline*}
The left member of the last equation is obeyed by $G'$, so that, the solvability condition is
\begin{multline*}\alpha_1\int_{-\infty}^{\infty}\zeta GG'dt=R\int_{-\infty}^{\infty}G^2(G')^2d\zeta\\
+\int_{-\infty}^{\infty}GG'
\left(1-\text{e}^{-\int_{-\infty}^\zeta(G^2-P_{th}^2)\Theta(G^2-P_\text{th}^2)d\zeta'}d\zeta\right),
\end{multline*}
which gives
\begin{widetext}
\begin{multline}
\alpha_1=-\frac{4}{15}RC^2+\frac{\text{e}^{-\sqrt{C-P_\text{th}^2}+P_\text{th}^2\zeta_1}}{\sqrt{C}}\int_{-\zeta_1}^{\zeta_1}GG'
\text{e}^{-\sqrt{C}\text{tanh}(\sqrt{C}\zeta')+P_\text{th}^2\zeta'}d\zeta'\\
+\frac{P_\text{th}^2}{2\sqrt{C}}\left(1-\text{e}
^{-2\sqrt{C-P_\text{th}^2}+2P_\text{th}^2\zeta_1}\right) \label{alpha1}
\end{multline}
\end{widetext}
where $\zeta'=\zeta-\zeta_0$ and $\zeta_1$ is the instant at which $G(\pm \zeta_1)=P_\text{th}$.
The integral in the last expression may be written in closed form as a series but here we solve it numerically.
Nevertheless, for $P_\text{th}=0$, the integral is easily solved analytically such that $\alpha_1$ simplifies to
\begin{equation}\alpha_1=-\frac{4}{15}RC^2-(C^{-1/2}-1)+(C^{-1/2}+1)\text{e}^{-2\sqrt{C}}.
\label{alpha1-Pth0}
\end{equation}
In the limit of small $C$, this equation reduces to
$\alpha_1=-\frac{4}{15}RC^2+\frac{2}{3}C$.
Moreover, a graphical inspection of expression (\ref{alpha1-Pth0}) shows that it is always below the curve $\frac{2}{3}C$ and in fact it tends to 1 as $C$ increases.
On the other hand,
whenever the peak intensity is close to the intensity threshold,
i.e., $\sqrt{C}\sim P_\text{th}$, we may expand the exponentials in
equation (\ref{alpha1}) up to first order and obtain
\begin{equation}
\alpha_1=-\frac{4}{15}RC^2+\frac{2}{3\sqrt{C}}(C-P_\text{th}^2)^{3/2}.
\label{alpha1-smallamp}
\end{equation}

Note that both this expression and the approximate expression for
the acceleration for small $C$ and $P_\text{th}=0$ exhibit $C^2$ and
$C$ dependencies which are associated, respectively, to Raman and
plasma effects. Such dependencies imply that the
plasma effect is expected to dominate for small peak amplitude
pulses, with the acceleration taking positive values which are
proportional to the square of the peak amplitude. Conversely, as the
peak amplitude increases, the soliton trajectory should be mainly
controlled by Raman effect, which leads to a negative acceleration
that is dependent on the forth power of the peak amplitude.

\section{\label{shooting}Pulse profiles and accelerations}
We then used a shooting method to obtain the pulse solutions of
equation (\ref{Peq}) and respective accelerations. For this purpose,
we first analyse the asymptotical form of this equation for pulse
solutions that vanish at the limits $\zeta\rightarrow\pm\infty$ which is
given by

 $$P''+(\alpha \zeta-C-\chi\Lambda_\infty)P=0,$$
 where
 $$\Lambda_\infty=\left\{\begin{array}{lcr} 0 & \text{if} & \zeta\rightarrow -\infty \\
 1-\text{e}^{-\int_{-\infty}^\infty \Delta {P^2}\Theta (\Delta {P^2})d\zeta'} & \text{if} & \zeta \rightarrow \infty\end{array}\right.
 $$
that, using $z=-\alpha^{1/3}\zeta+\alpha^{-2/3}(C+\chi\Lambda_\infty)$, may be transformed to an Airy equation:
$$P''-zP=0.$$
This result anticipates that the pulse solutions with $\alpha>0$
have tails that are exponentially decreasing as $\zeta\rightarrow
-\infty$ as the Airy solution Ai$(z)$ for $z\rightarrow \infty$ and
may have tails that are also exponentially decreasing as $\zeta$
increases  as the solution Bi$(z)$ for $z\rightarrow 0^+$ but that
eventually exhibit Airy oscillations (even if very, very small). For
$\alpha<0$, the contrary is true.

Considering those asymptotic behavior, we designed our shooting procedure as follows. First, we have fixed the
acceleration $\alpha$ and, starting from the left tail, we
integrated forward using initial conditions that conform with the
corresponding Ai$(z)$. The actual location of the pulse in the
$\zeta$ axis may be estimated using the perturbation approach
described in the previous section but, since it is not very far from
$\zeta=0$, the first estimate for the left tail location
$\zeta_\text{minus}$ may be obtained as if $P(\zeta)\sim G(\zeta)$.
Then, the shooting procedure checks if $P$ and $P'$ are already very
small at some point in the right tail, and improve the starting
$\zeta$ in order to obtain the actual pulse profile for the chosen
acceleration. Therefore, this procedure allow us to establish the
relationship between the acceleration and the pulse characteristics,
namely, its peak amplitude.

\begin{figure}
\centering\includegraphics[width=7.5cm]{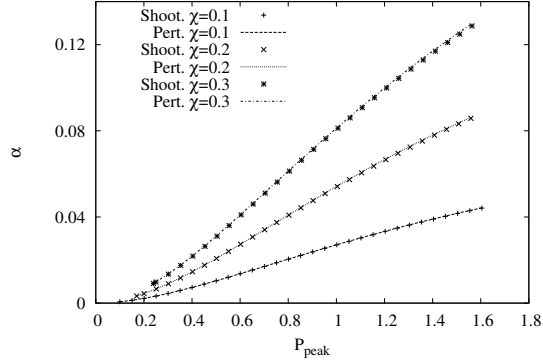}
\caption{\label{aceleration}Dependence of acceleration parameter
$\alpha$ on the pulse peak amplitude for $P_\text{th}=0$ and for three
different $\chi$ values. Points are shooting results and lines are
for the perturbation expression.}
\end{figure}

\begin{figure}
\subfigure[]{\label{P0p3}\centering\includegraphics[width=7.5cm]{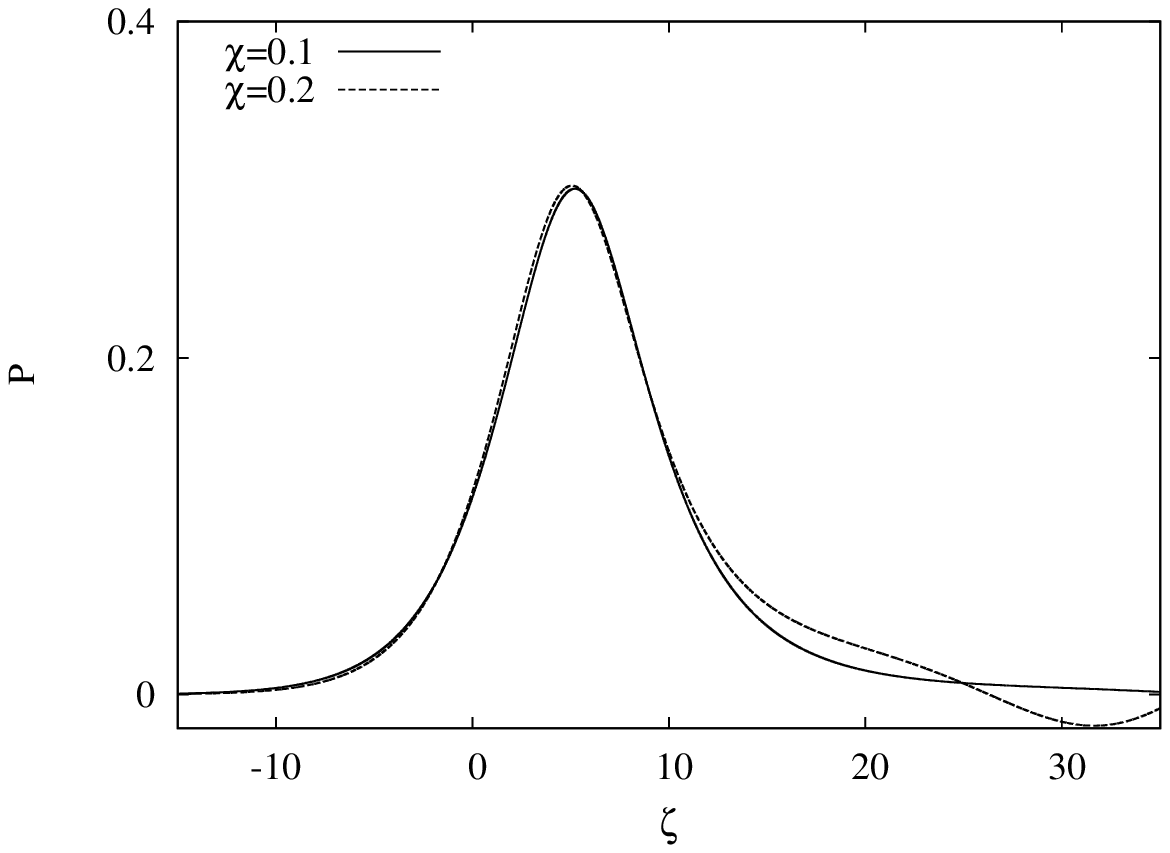}}
\subfigure[]{\label{P0p8}\centering\includegraphics[width=7.5cm]{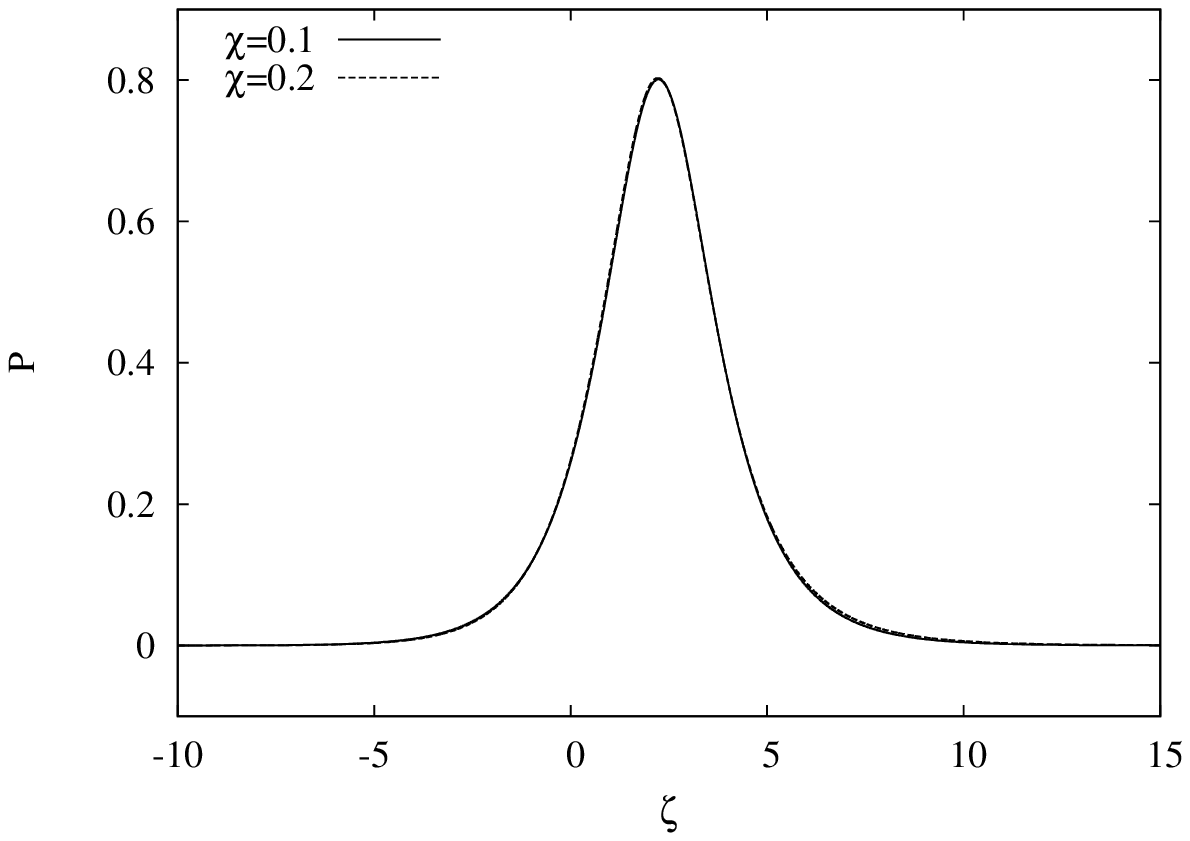}}
\caption{\label{profiles}Pulse profiles with peak amplitude close to (a) 0.3 and (b) 0.8 for two different $\chi$ values.}
\end{figure}

Our results show that, as long as $\chi$ is small, the dependence of
the acceleration on the peak amplitude is in fact very similar to
the one obtained by perturbation, that is, by using expressions
(\ref{alpha1}) or (\ref{alpha1-Pth0}) with $\sqrt{C}$ replaced by $P_\text{peak}$.
Let us first discuss the results for $P_\text{th}=0$. Figure
\ref{aceleration} shows the acceleration as function of peak
amplitude of $P(\zeta)$ for three different strengths of the
plasma term and without the Raman term. As shown, the
acceleration increases with the amplitude of the peak and a good
agreement exists between the shooting results and the perturbation
expression (\ref{alpha1-Pth0}). Nevertheless, for $\chi=0.3$ there is an
observable difference between the two results that can be attributed
to the deviation of the pulse profile from the sech shape considered
in equation (\ref{alpha1-Pth0}). The absence of results for low
$P_\text{peak}$ in the curves for $\chi=0.2$ and $\chi=0.3$ is due
to the inability of our shooting to find a solution with a small
right tail. In fact, the pulse profiles with small amplitude are
located close to the zero of the Airy $z$ axis, effect that is more
pronounced as $\chi$ grows. This means that, in those situations, the
solution is no longer similar to a sech profile but it has long, and
eventually oscillatory, tails. Figure \ref{profiles} presents two
sets of solutions for $P_\text{peak}\sim 0.3$ and $P_\text{peak}\sim
0.8$. The first set shows considerable differences at the right
tail, with the pulse for $\chi=0.2$ having a longer tail. In the
second set, the shape differences are not so evident since for this
peak amplitude, both solutions are already similar to each other and
with the sech shape.

\begin{figure}
\centering\includegraphics[width=7.5cm]{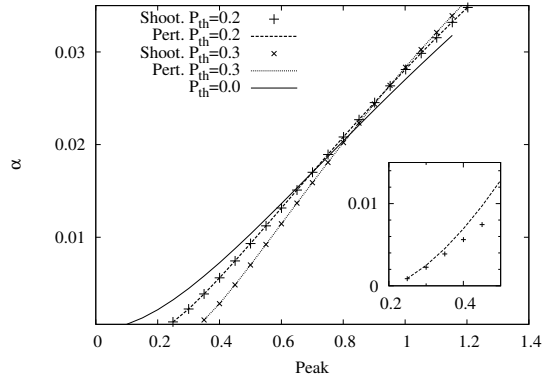}
\caption{\label{aceleration2}Dependence of acceleration parameter
$\alpha$ on the pulse peak amplitude for different $P_\text{th}$
values and fixed $\chi=0.1$. Points are shooting results and lines
are for the perturbation expression (\ref{alpha1}). The inset shows the shooting
accelerations for $P_\text{th}=0.2$ compared with $\alpha$ computed
using (\ref{alpha1-smallamp}) with $\sqrt{C}$ replaced by
$P_\text{peak}$.}
\end{figure}

\begin{figure}
\subfigure[]{\label{P0p2Pth}\centering\includegraphics[width=7.5cm]{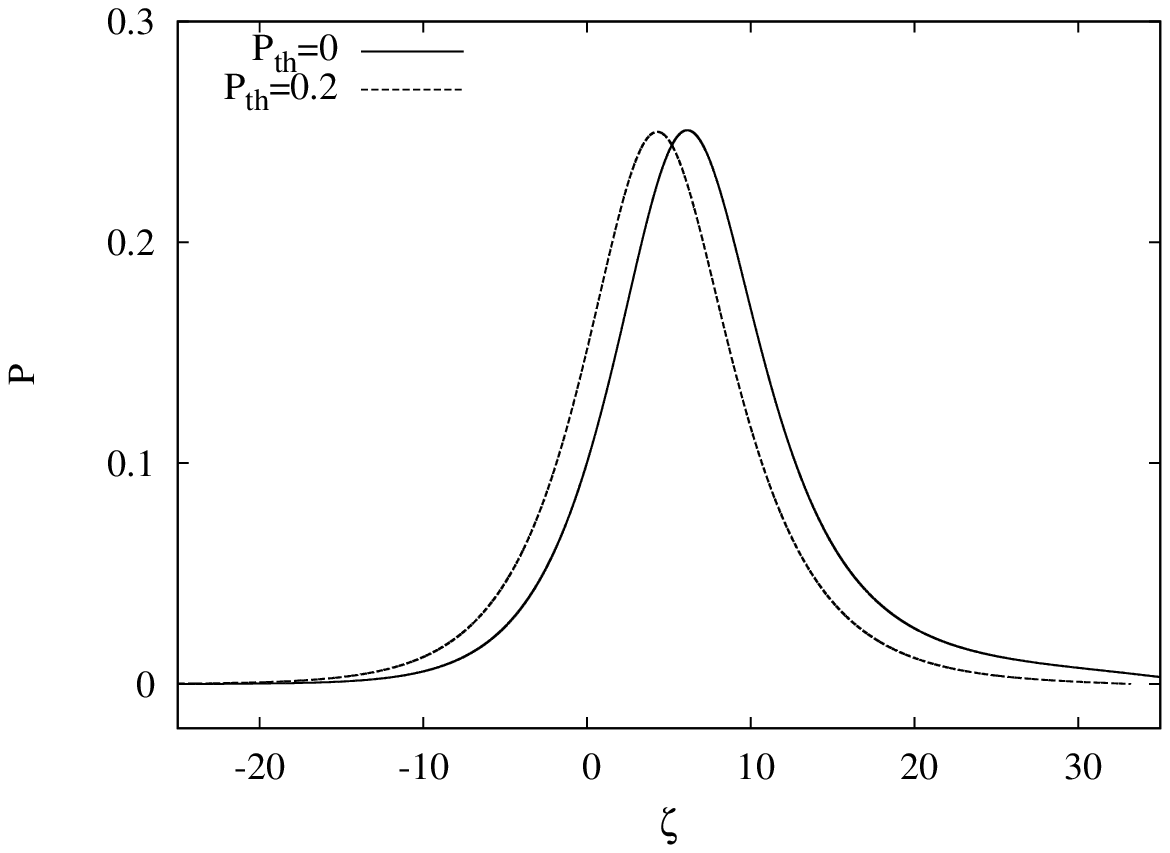}}
\subfigure[]{\label{P0p5Pth}\centering\includegraphics[width=7.5cm]{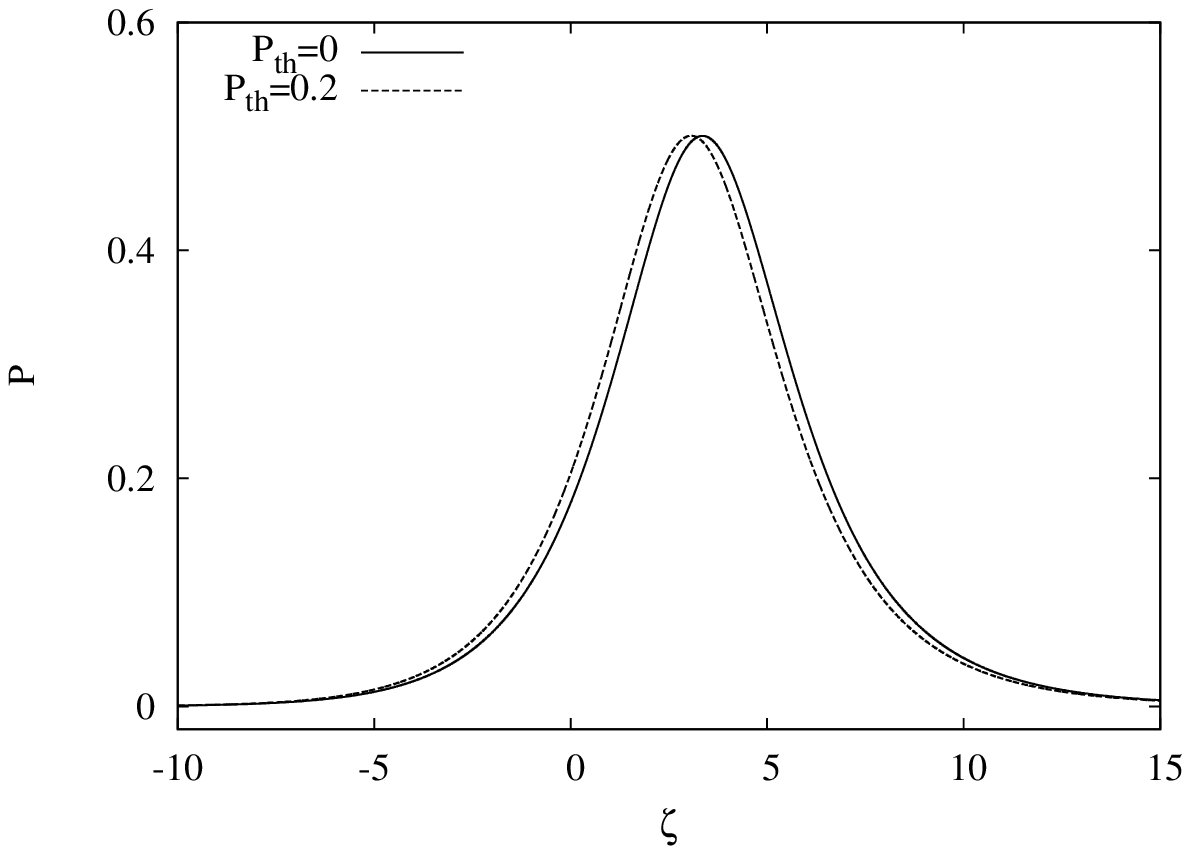}}
\caption{\label{profilesPth}Pulse profiles with peak amplitude close
to (a) 0.25 and (b) 0.5 for two different $P_\text{th}$ values and
fixed $\chi=0.1$.}
\end{figure}

Concerning the numerical results for $P_\text{th}\neq 0$, the
accelerations are again in good agreement with the perturbation
expression if $\chi$ is small. As figure \ref{aceleration2} shows,
when compared with the results for $P_\text{th}=0$, the
accelerations are lower for small peak amplitudes and larger
otherwise. Furthermore, the acceleration decreases with increasing
$P_\text{th}$ for smaller peak amplitudes and the inverse is true
for larger peak amplitudes. Also represented in this figure is the
acceleration resulting from the approximate expression
(\ref{alpha1-smallamp}) with $\sqrt{C}$ replaced by $P_\text{peak}$,
which, as anticipated, is close to the shooting results when
$P_\text{peak}\sim P_\text{th}$. The pulse shape differs from the
sech also for some peak amplitudes close to $P_\text{th}$ but this effect is smaller as
$P_\text{th}$ increases. Note that as $P_\text{th}$ increases, the
possible peak amplitudes that make the plasma term nonzero
are also increasing, since the latter should be larger than the
first. Figure \ref{profilesPth} compares pulse profiles for
$P_\text{th}=0$ and 0.2.

\begin{figure}
\centering\includegraphics[width=7.5cm]{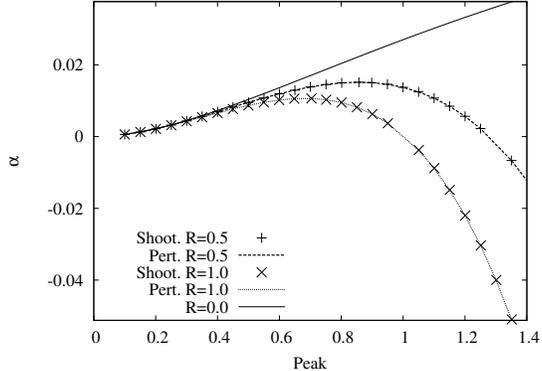}
\caption{\label{aceleration3}Dependence of acceleration parameter
$\alpha$ on the pulse peak amplitude for different $R$ values and
fixed $\chi=0.1$ and $P_\text{th}=0$. Points are shooting results
and lines are for the perturbation expression.}
\end{figure}

As we introduce $R$ different from zero, i.e., include the Raman
term, the behavior of the acceleration with the peak amplitude is
illustrated  in Figure \ref{aceleration3}. For small peak
amplitudes, $\alpha$ increases with the peak amplitude whose
behavior is characteristic of the plasma effect. However,
as the peak amplitude increases further, the acceleration starts to
decrease into the region of negative accelerations that are
characteristic of the accelerating solitons of the NLS plus IRS
\cite{facao10}. Note that, similarly to what happened when only
the plasma term was present, the shooting was performed forward, but in the cases of negative $\alpha$,
the estimates of $P$ and $P'$ in the left tail were taken from Bi.
Still for the case $R\neq 0$, the proximity of the pulses to the
$z=0$ in the Airy axis for smaller peak amplitudes is lost as we
introduce the Raman term. However, as the peak amplitudes increase
to values for which the acceleration is negative and large in
modulus, the pulse returns to the neighborhood of the Airy $z=0$ and
starts to develop long tails but, in this case,  to the left.

\begin{figure}
\centering\includegraphics[width=7.5cm]{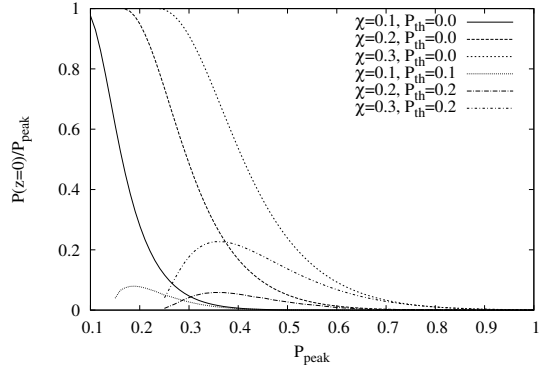}
\caption{\label{Pz0} Dependence of the relative amplitude at $z=0$
on the pulse peak amplitude for different $\chi$ and $P_\text{th}$
values.}
\end{figure}

\begin{figure}
\subfigure[]{\label{nonl0p5}\centering\includegraphics[width=7.5cm]{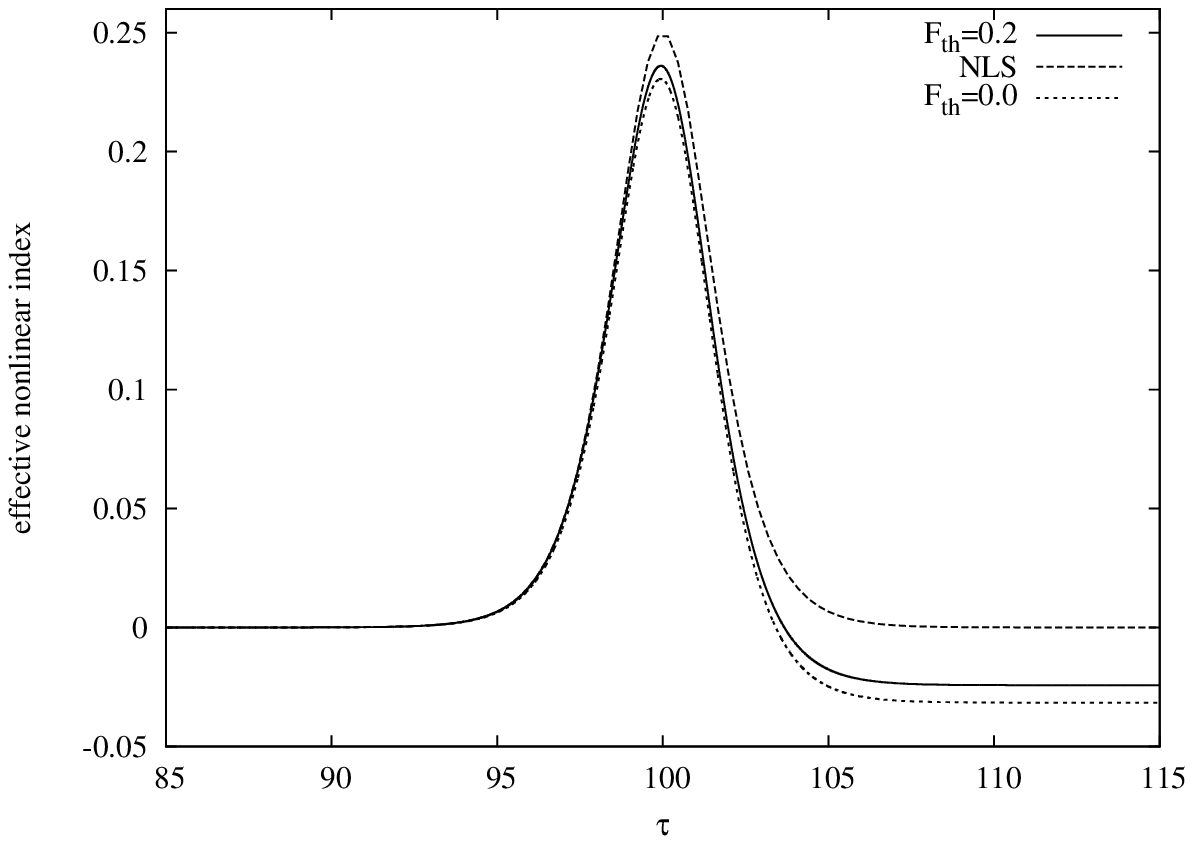}}
\subfigure[]{\label{nonl0p1}\centering\includegraphics[width=7.5cm]{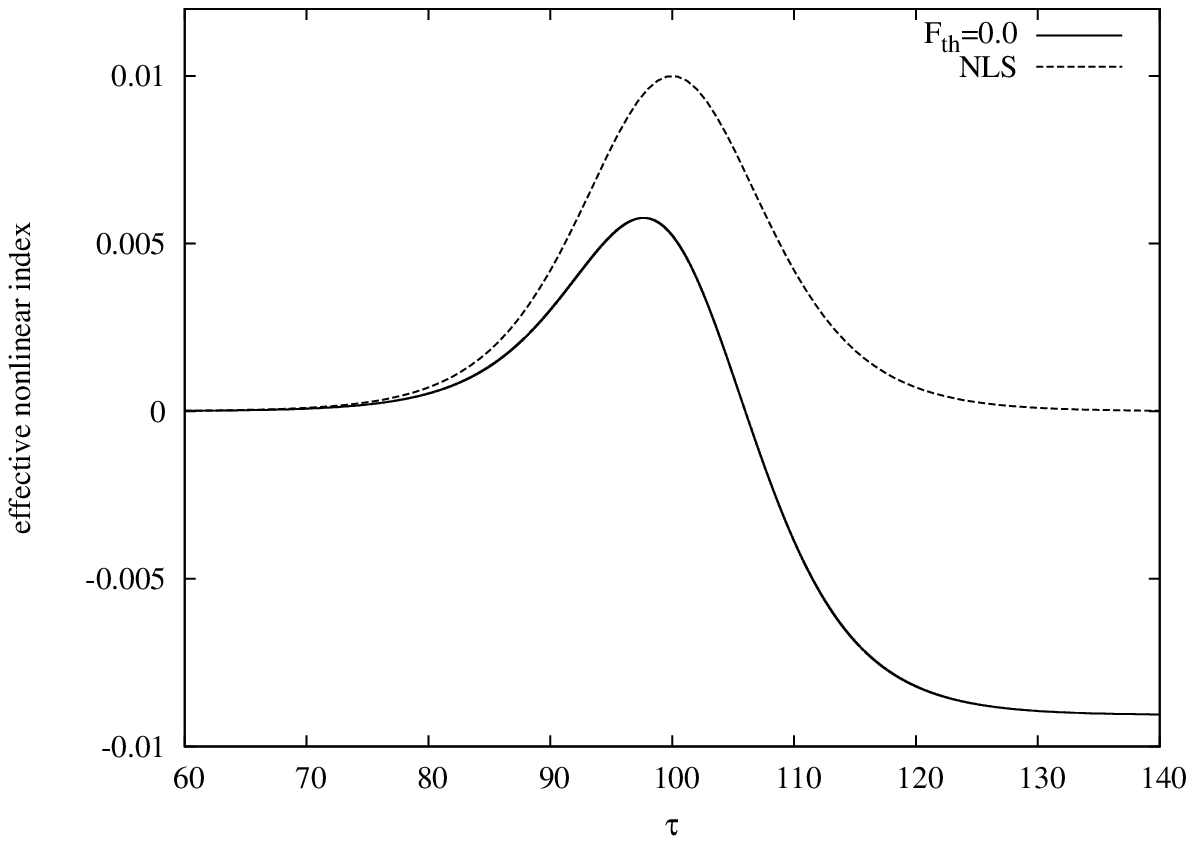}}
\caption{\label{nonlinear}Effective nonlinear index for peak
amplitudes close to (a) 0.5 and (b) 0.1 for fixed $\chi=0.1$.
Comparison with the NLS case.}
\end{figure}

Let us return to the case $R=0$. Our results indicate that the pulse profiles develop long tails
whenever the peak amplitude is small. Since these long tails can be
associated with the oscillatory behavior of the Airy functions for
negative $z$, we plotted in figure \ref{Pz0} the relative pulse
amplitude at $z=0$ as a function of the peak amplitude for different
values of $\chi$ and $P_\text{th}$. As expected, this relative
amplitude increases with $\chi$. On the other hand, when
$P_\text{th}=0$, this relative amplitude increases with the decrease
of $P_\text{peak}$ but, for $P_\text{th}\neq 0$ the curves exhibit a
maximum for a given value of $P_\text{peak}$ that is not large when
compared to $P_\text{th}$. The existence of the long tails for small
peak amplitude pulses is not readily understood since for those
amplitudes the plasma term is smaller. Also, we know
that in the presence of IRS the pulses develop long tails if the
Raman term is large, which happens for large $\tau_R$ or short
pulses (large peak amplitudes). In order to better understand the
deviation from the sech shape in the presence of the plasma
term, we plotted the effective nonlinear refractive index
$|q|^2-\tau_R(|q|^2)_\tau-\phi_T\left(1-\text{e}^{-\sigma\int_{-\infty}^\tau\Delta|q|^2\Theta(\Delta|q|^2)d\tau'}\right)$
for several cases. Figure \ref{nonlinear} compares two of those
cases against the typical nonlinear refractive index of the NLS. We
may observe that, although the magnitude of the effect of the plasma
term is greater for $F_\text{peak}=0.5$, the relative deviation from
the NLS case is larger in the $F_\text{peak}=0.1$ case. There, we
may also see that the introduction of nonzero $P_\text{th}$
decreases the plasma effect what was fully expected since this means
that only part of the pulse creates the plasma.

\section{Direct numerical integration}
\begin{figure}
\subfigure{\label{map-Pth0p2}\centering\includegraphics[width=7.5cm]{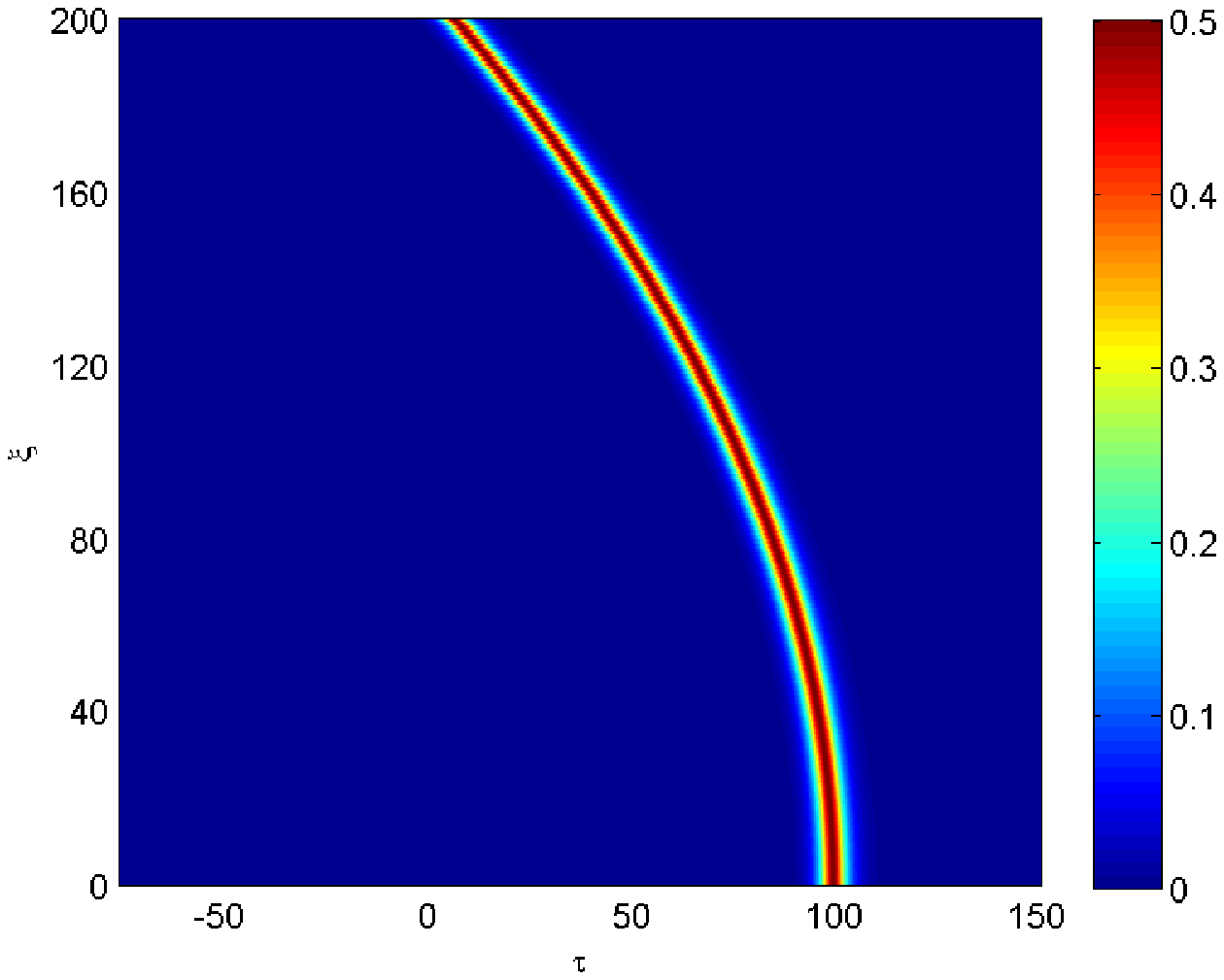}}
\subfigure{\label{map-chi0p2}\centering\includegraphics[width=7.5cm]{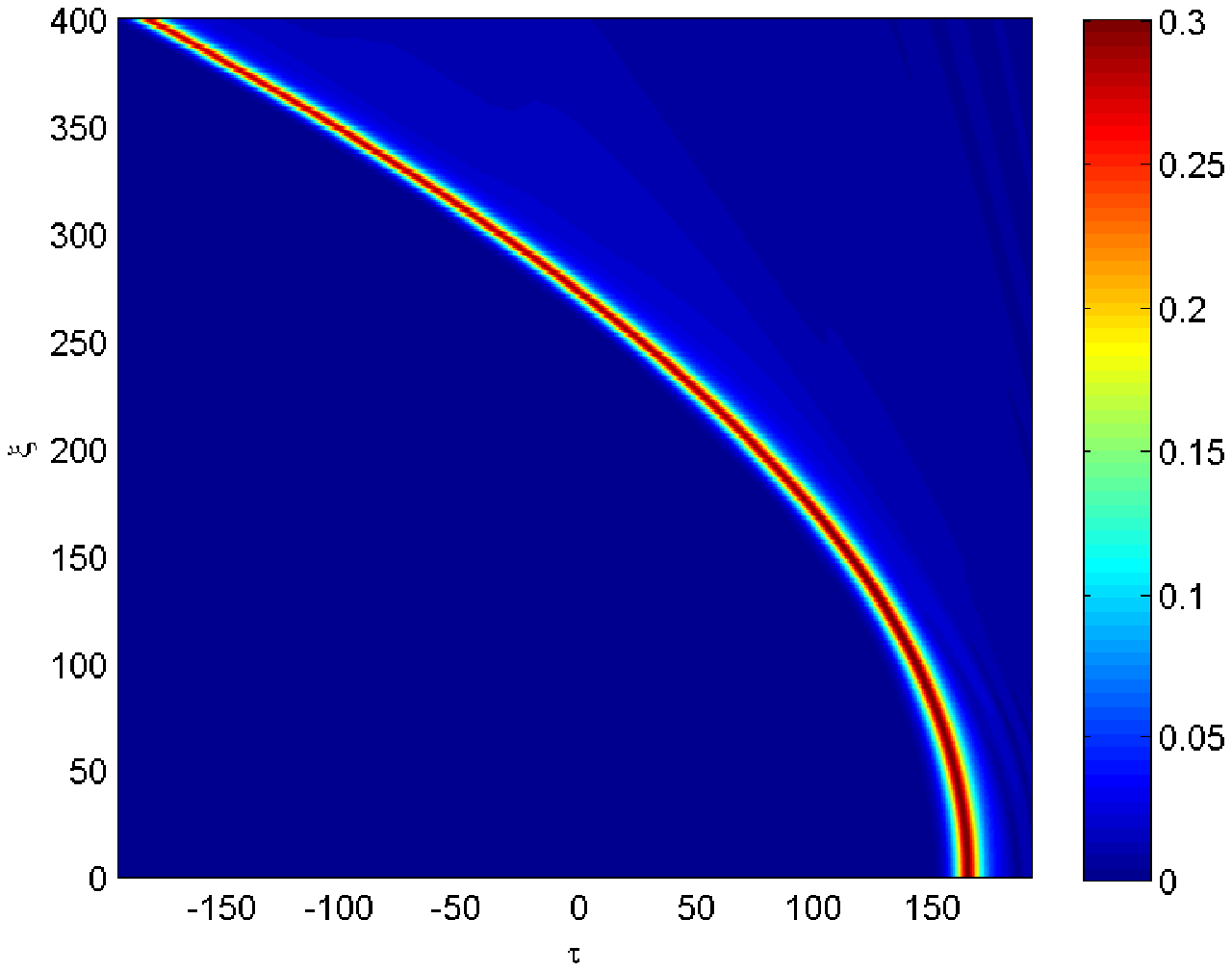}}
\caption{Evolution of pulse solutions, $|q|$  with (a)
peak amplitude equal to 0.5 for $\chi=0.1$ and $P_\text{th}=0.2$ and (b) peak amplitude equal to 0.3 for $\chi=0.2$ and $P_\text{th}=0$. We
have used $\sigma=1$.}
\end{figure}

\begin{figure}
\centering\includegraphics[width=7.5cm]{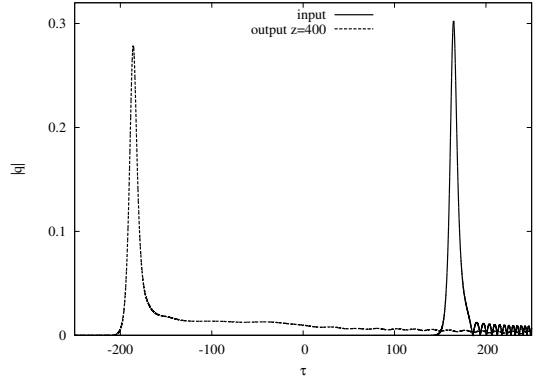}
\caption{\label{in-out}Input and output for the simulation whose
contour is on Fig.~\ref{map-chi0p2}.}
\end{figure}

Direct numerical simulations of the full equation (\ref{pde}) were
performed using pseudospectral codes in order to study the stability
of the solutions described in the last section and to confirm
accelerations and existence of tails. In general, the solutions
found by the shooting procedure are stable and evolve along the
predicted trajectory. Figure \ref{map-Pth0p2} is a contour graph
showing the evolution of the pulse profile of Fig.~\ref{P0p5Pth} for
$P_\text{th}=0.2$. The trajectory is in full agreement with the
predicted acceleration.
Whenever long tails were found in the shooting procedure, they were
confirmed in the propagating solution. In cases of very long tails,
the solution is no longer stable but decays. Figures \ref{map-chi0p2} and \ref{in-out} show the evolution and output
as obtained for the pulse profile of Fig.~\ref{P0p3} corresponding to $\chi=0.2$.
The same kind of behavior was already observed with the Raman
accelerating solutions \cite{facao10} and is consistent with infinite
energy of the Airy solutions Ai$(z)$ and Bi$(z)$.
As discussed in section \ref{shooting}, pulse solutions in the positive Airy axis and far away from its
zero, have exponential decay in both tails, similar to Ai to the
left and to Bi to the right (the inverse happens for $\alpha<0$).
The algebraically decaying oscillations of Bi$(z)$ for negative $z$
will only occur far way in the right tails (left tails for
$\alpha<0$). However, if the solutions are in the positive Airy axis
but close to the zero one of the tails will behave like a
combination of Ai and Bi, it will exhibit the typical algebraically
decay and it will shed radiation away during propagation. Figures
\ref{map-chi0p2} and \ref{in-out} report these latter behavior.


Finally, let us return to the physical variables and calculate the actual acceleration and frequency shift. The adimensional acceleration $a$ does not correspond directly to an acceleration in physical units, nevertheless let us define the acceleration $a_r$ as the second derivative of the temporal peak position in the group velocity reference frame relatively to the propagation distance $z$, namely
\begin{widetext}
\begin{equation*}
a_r =\frac{d^2 t_\text{peak}}{dz^2}=\frac{|\beta_2|^2}{t_0^3}\frac{d^2
\tau_\text{peak}}{d\xi^2}=-\frac{|\beta_2|^2}{t_0^3}\frac{a}{2}=-\frac{A_\text{eff}^3\gamma^3}{2\tilde{\sigma}^3|\beta_2|}\alpha(\chi,P_\text{th},
P_\text{peak})
\end{equation*}
\end{widetext}
Note that the negative signal only implies that the pulse is traveling toward negative $t$ but since $t$ is measured in a reference frame that travels with the group velocity for $\omega_0$, the pulse is gaining velocity whenever this acceleration is negative.
This change in velocity is due to a deviation in frequency that is linear with the distance $z$, as expressed in the phase (\ref{phase}), and given by
$$\Delta\omega=-\frac{d\theta}{dt}=-\frac{a_r}{|\beta_2|} z$$
Since equation (\ref{pde}) neglects the photoionization related
losses that are small for pulses whose peak amplitude is comparable
with the threshold, for $\chi$ not too large, we may use expression (\ref{alpha1-smallamp}) for $\alpha$ and approximate the frequency approximated by
$$\Delta\omega=-\frac{8}{15}\frac{t_R\gamma^2}{|\beta_2|}\psi_\text{peak}^4z+\frac{k_0(\omega_T/\omega_0)^2\tilde{\sigma}}{3A_\text{eff}}
\frac{\left(\psi_\text{peak}^2-\psi_\text{th}^2\right)^{3/2}}{\psi_\text{peak}}z$$
which gives the standard result for the IRS \cite{gordon86,facao10} and the effect of plasma growing with order $\psi_\text{peak}^2$.

\section{Conclusions}
We have found the self-similar accelerating solutions of a  generalized NLS that includes IRS and a term for plasma induced nonlinearity. This equation models the propagation of pulses in gas filled HC-PCFs where it has occurred photoionization of the gas. The solutions are very close to the NLS sech soliton as long as the strength of the plasma term is relatively low and the solution amplitude is relatively large. The accelerations and the blueshifting increase with the peak amplitude of the pulses. In case of pulse solutions whose peak intensity is close to the photoionization threshold, which are the ones for which the equation better models the physical effects, the frequency blueshift increases in the same order as the square of peak amplitude. However, also the same solutions, whose peak amplitudes are close to the threshold, may exhibit a profile that is considerably different from the sech, have long tails and decay along the propagation distance.

\acknowledgments{This work was partially supported by the Funda\c{c}\~{a}o para a Ci\^{e}ncia e Tecnologia, FCT, and European Union FEDER program and PTDC programs, through the projects
PTDC/EEA-TEL/105254/2008 (OSP-HNLF),
PTDC/FIS/112624/2009 (CONLUZ) and PEst-C/CTM/LA0025/2011.}

\end{document}